# *Characterizing IOMT/Personal Area Networks Landscape*

**Adil Rajput**

(aiilahi@effatuniversity.edu.sa)

*Assistant Professor, Information System Department, Effat University*

*An Nazlah Al Yamaniyyah, Jeddah 22332, Jeddah, Saudi Arabia*

**Tayeb Brahimi**

(tbrahimi@effatuniversity.edu.sa)

*Assistant Professor, Natural Sciences, Mathematics, and Technology Unit, Effat University*

*An Nazlah Al Yamaniyyah, Jeddah 22332, Jeddah, Saudi Arabia*

## Abstract

The attention garnered by the Internet of Things (IoT) term has taken the world by a storm. IOT concepts have been applied to various domains revolutionizing the way business processes are conducted. Health Informatics is a nascent multidisciplinary field that aims at applying information engineering concepts to healthcare. The information traditionally came from a variety of sources such as healthcare IT systems but recently is being stored in variety of IOT devices. Application of IOT concepts is becoming the norm giving rise to the Internet of Medical Things (IOMT). However, IOMT introduces many challenges such as the real-time nature, security, and privacy of data along with others. Thus, understanding the underlying structure of IOMT is of paramount importance. Given the importance of the underlying networks, we explore various solutions currently employed in developing the Personal Area Networks (PANs) that are the underlying cornerstone of IOMT). PANs allow the sensors to measure the change in various stimuli and transmit the information via LANs and WANs to various stakeholders. This chapter will look at three standards that are currently used by both academia and industry along with Body Area Networks (BANs). The chapter also provides a survey of prevalent IOMT applications along with various vendors that provide such services.

# 1 Introduction

Advances in the communication and technology realm have given rise to an interconnected world which has eliminated the need for human intervention. While the fields of Artificial Intelligence (AI), Decision Support System (DSS), and Machine Learning (ML) have been advocating various techniques to enhance the decision making and more importantly reduce the human-related errors, it is the inexpensive price of hardware that has made the IoT (Internet of Things) a reality. The term "Internet of Things" was first coined by Ashton (Ashton, 2009). Although many definitions exist for IoT, the IoT environment is characterized by a collection of "devices" connected together via a "network". Such devices need both a hardware and software component - a system known as Cyber-Physical System (CPS).

The term CPS refers to the intertwined system of hardware and software components that allows the software to control the hardware components minimizing or eliminating the use of the human intervention. The prerequisite for such systems to operate is the underlying ubiquitous network be it on a Local Area Network (LAN) or Wide Area Network (WAN). Such a network allows the various hardware components to communicate with the software component in real time (or near real time). Almost in all cases, the CPS systems strive to minimize or eliminate the human intervention thus increasing the demand on both CPU, memory and the network. Given that the size of the hardware components is usually small (e.g., a wearable device), the power and memory capacity cause serious challenges as the size of the components connected and the demands placed on such devices.

## 1.1 IOMT and Health Informatics

The Internet of Medical Things (IoMT) - known as the healthcare IoT (Ashton, 2009) technology - ensures the availability and the analysis of healthcare data through smart medical devices and the web (Joyia et al., 2016; Dastjerdi et al., 2016; Islam et al., 2015, Singh et a;., 2014). Internet of Medical Things builds upon the underlying architecture of the IoT as it has a multitude of interconnected devices sending data over a network (mostly the Internet) to the cloud that can be accessed by the medical practitioners (Deliotte, 2018; Gatouillat et al., 2018, Díaz et al., 2016). Applications range from gathering data with remote monitoring systems to be analyzed in non-real time (e.g., sleep monitoring) to emergency notification services (e.g., pacemaker monitoring).

Devices such as Fitbit, Garmin, Xiaomi, or Misfit wireless fitness tracker are used to monitor patients' vital signs. Given the battery and computation limitations, the majority of the data is transmitted via the network (LAN and possibly WAN) placing high requirements on the network. Furthermore, the real-time nature of IoMT coupled with the nascent stage of development has brought forth various challenges to the helm. Such challenges include (but not limited to) security/privacy of data, efficient data handling, and transmission, and massive data volume (Joyia et al., 2016). Recent research and developments in advanced sensors, mobile applications, artificial intelligence, big data, 3D printing, and mobility have created new opportunities for medical technology companies to design and manufacture a wide range of affordable smart medical devices (Dey et al., 2018; Deliotte, 2018, McDonald et al., 2018; Raghupathi et al., 2014).

**1.2 Personal Are Networks**

The idea behind IOMT centers around the concept of Personal Area Networks (PANs). A PAN is a small network that centers around a sensor device (could be attached to a machine or a human). The basic premise behind PAN network is the following:

1. Various devices and/or human body will have sensors connected to them
2. Such devices communicate the status of each device using a network
3. The network formed is ad-hoc and limited in range
4. The devices have limited battery power and hence can be limited in terms of functionality
5. Given the low power limitations, the network throughput can be low.
6. Depending on the nature of the information, the information needs to be secured

The goal of this chapter is to describe the various components that form the underlying basis of Personal Area Networks. The crux of this chapter is presented in the next section. First, we describe the two main components of IoT/IoMT namely the physical and network components. Then we delve deeper into the network component and explain the three standards that underly the network component of the PANs along with an overview of Body Area Networks (BANs). Section 3 discusses some of the prevalent applications in the IoMT world. Section 4 offers the concluding remarks along with future research directions.

## 2 Architectural Landscape

Revisiting the Cyber-Physical system definition highlights the intertwined nature of the hardware and the software. Furthermore, the physical world we live in deals with information that is represented in an analog fashion. Such information is encoded as part of waves that can take on values over a continuum. On the other hand, computers and software only deal with a discrete set of values that can be mapped onto 0s and 1s. Both the health-care professionals along with the technology architects need to fully comprehend the underlying makeup of such systems to better address the implementation details along with the associated risks. Therefore, to delve into such details, we need to address the following:

1. Physical Components
2. Network Components

The Physical components allow the analog signals to be transferred from a device/body to a node that can process such information and receive instructions on steps to be performed. The Network components actually allow the messages to be transferred between devices – both in asynchronous and synchronous mode.

## 2.1 Physical Components

The IoT paradigm is made possible by both a physical and network component. The physical components make it possible for the devices to both follow changes on the physical level and take an action.

The physical level is composed of various Micro-Electronic-Mechanical Systems (MEMS) devices (Hesu, 2002). In simple terms, MEMS systems are an amalgamation of both micro-electronic and micro-mechanical components along with a micro-sensor and a micro-actuator referred to as sensors and actuators going forward. The success of MEMS along with low cost of hardware has also given rise to Nano-Electronic-Mechanical Systems (NEMS) which follows the same idea but the size of hardware is even smaller than MEMS. While the electronic and mechanical components are worked on by a small number of experts, it is the sensors and actuators components that practitioners in the IoT realm deal with.

### 2.1.1 Physical Components

The sensors provide the ability to monitor physical stimuli such as temperature or humidity. This part is the job of the microelectronics components. Based on the certain changes in the stimuli, the actuators take certain actions underpinned by the micromechanical system. As an example, consider the time-temperature monitoring system which after detecting the temperature dropping past a certain temperature threshold would cause the heat to turn on. While such abilities used to be hardcoded into the hardware, the MEMS systems are characterized by a Central Processing Unit that would make programming such devices a reality.

The sensors can be both digital and analog. A digital signal has a finite set of values while an analog signal is a continuous change of a given parameter over a period of time - mostly voltage. The voltage can oscillate between a minimum and maximum range on a continuous basis. As an example, some digital sensors represent the 'on'/ 'off' state by using the value of 3.3 Volts. In other words, the sensor can take on only two values (0 for off or 3.3 V). On the other hand, an analog sensor can take any value on the continuum between 0 and 3.3V. The analog sensors need to provide a way to convert the analog signal to a digital one so that it can be processed by a software.

An actuator is a mechanical component that is responsible for performing a certain action based on the input from the sensor. The actuator needs a source of energy that can be provided via various means such as an electric signal, liquid compression, change in pressure etc. In all cases the energy is received as a control signal which will then be converted into mechanical form causing the action to be performed.

## 2.2   Network Component

The backbone of an IoT infrastructure is the network component. While a wired network can be used, wireless networks are now considered the norm. The hardware components describe in section 2.1 are connected via a network - hence the term Cyber-Physical systems. The term Wireless Sensor Networks (WSN) has become synonymous with the network component of the IoT systems. However, Personal Area Networks (PAN) more accurately describe the concept at hand.

While a LAN allows connectivity with a high rate of transmission, PANs specifically cater to devices that have limited power and bandwidth. While the narrow definition assumes that the components belong to a particular individual, in reality, PANs take on a bigger meaning e.g., when applied to industrial IoT. Recall that IoT is characterized by devices that are low on power and resources. The challenge, therefore, becomes as to how information can be transmitted with such constraints.

The IEEE 802.15 working group deals with defining the standards for wireless PANs. While the working group has defined ten areas for research, we will discuss the following four standards as they are deemed essential for transmitting the data to other devices.

1. Bluetooth
2. IEEE 802.1.5.4 Low rate WPAN
3. IEEE 802.15.3 High rate WPAN
4. IEEE 802.15. 6 Body Area Networks

### 2.2.1 Bluetooth

Historically the Bluetooth standard was termed as 802.15.1 but is no longer maintained by IEEE. Rather the standard is now maintained by the Bluetooth SIG (https://www.bluetooth.com/specifications/bluetooth-core-specification). The Bluetooth technology is encapsulated in a chip that can be part of any particular device. The technology uses the master/slave concept which is referred to as host and controller. A Bluetooth host receives information from the device using a cable (or simulation of a cable) that is submitted to another mobile Bluetooth device using a special frequency.

#### 2.2.1.1 Protocol Stack

Consider a typical scenario for a Bluetooth application. A host device such as an automobile music system has the Bluetooth capability turned on. A user wants to user her smartphone to play an audio clip. The smartphone device is paired with the music system which allows the synchronous playing of the audio clip on the car's music system allowing the car system to control the smartphone device. What appears to be a seamless operation for an end user requires sending and receiving traffic over a set of protocols. The layers described in Figure 1 present a typical Bluetooth Operation (Please note that the list is not exhaustive).

The radio is responsible for receiving and transmitting the actual signal using a given radio frequency. The Link Management Protocol (LMP) protocol (Lang, 2005) is mostly responsible for the connection at this level. The Low Energy Link Layer (Akyildiz et. al., 2002) is also employed to minimize energy usage. The Baseband technology converts analog signals to digital signals that can be transmitted using various protocols.

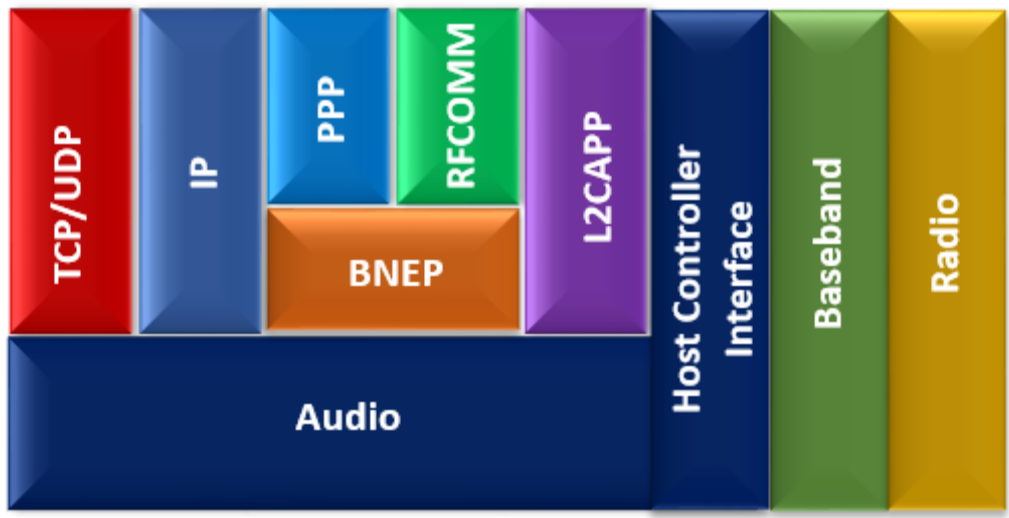

**Figure 1**  Bluetooth Protocol Stack

The Baseband technology, in turn, is divided into a physical and logical part. The physical channel is the actual transmission of radio signals over a given frequency. Given the probability of clashes on such frequency, the physical channel has the ability to switch the frequencies over a given range. In addition, the physical layer can be either a Synchronous Connection-Oriented (SCO) or Asynchronous Connection-Less (ACL). As suggested by the names SCO is used for appoint-to-point applications such as voice applications. ACL, on the other hand, allows a publish-subscribe model implementation between a master and one or more slaves.

The Bluetooth has been divided into five logical channels.  Such logical channels transfer different types of information. LC (Control Channel) and LM (Link Manager) channels are employed in a synchronous fashion. The UA, UI and US channels are used to carry asynchronous information. All these channels happen at the Link level. The Link Manager transmits the signal to the Host Controller Interface (HCI) that is also responsible for setting various parameters such as security etc. Once the signal is passed to the HCI, it deals with the Logical Link Control and Adaptation Protocol (L2CAP) (Kardach, 2000) whose sole purpose is to pass on the information from the Baseband Layers to higher protocols. In simpler terms, L2CAP serves a similar purpose to IP protocol when it comes to basic Network connectivity as we know it. The L2CAP layer allows various functions such as reassembly of packets, allow for synchronous and asynchronous communication etc. The importance of L2CAP mandates a detailed discussion which is beyond the scope of this chapter. The authors highly encourage users to explore this further to understand the intricacies of Bluetooth technology.

The L2CAP packets can be delivered between two devices using Point-to-Point (PPP) protocol or in an asynchronous way. The RFCOMM protocol provides the serial line interface and hence the packets are transferred using a typical IP network. The Bluetooth Network Encapsulation Protocol (BNEP) on the other hand is used to encapsulate L2CAP packets that can be transferred over a typical IP network. Lastly, the Service Delivery Protocol (SDP) allows various devices to negotiate a connection at the L2CAP layer as shown in figure 1.

### 2.2.1.2 Pico and Scatter Networks

Now that we have a basic comprehension of the various protocols that come into play, let us briefly discuss the types of networks Bluetooth technology allows. Recall that the WPAN formed using a Bluetooth technology is limited in distance in terms of range. We alluded to the concept of Master/Slave in the previous section but we will explain the concept further in the context of Bluetooth technology. The Master device relates to the host that is responsible for transmitting and receiving information from various devices – termed as slaves.

A Pico network is the basic unit in a Bluetooth network where a master device allows/ requests a connection from various slave devices. The ad-hoc network formed once the devices come into contact is known as a Pico Network, Figure 2 depicts such a network (PicoNet). The concept is similar to the formation of a LAN/WLAN where many devices are connected to one switch/router

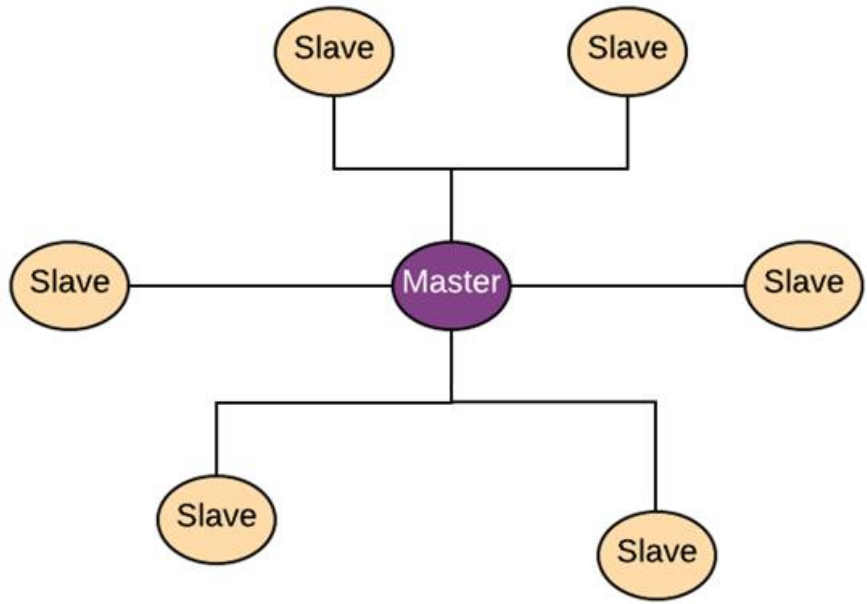

**Figure 2** PicoNet

The combination of two or Pico networks, shown in Figure 3, is called a Scatter Network. Note that a device can act as a slave in one Pico network while serving as a master in another one. This flexibility allows efficient dissemination of information using a Bluetooth network.

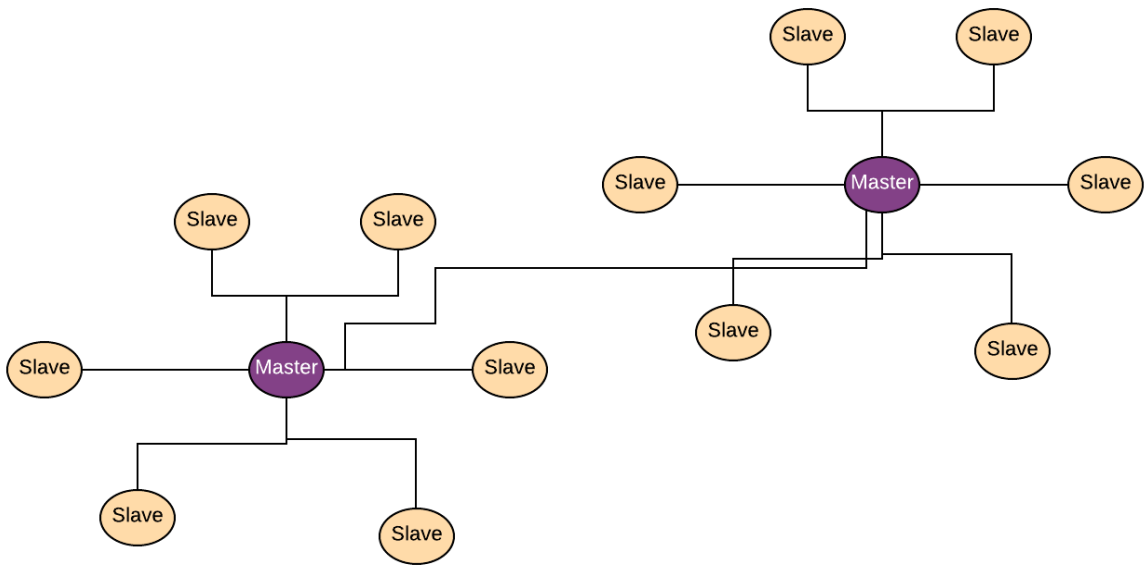

**Figure 3** ScatterNet

## 2.2.2 Law Rate WPAN

A Low Rate WPAN (LR-WPAN) provides a cost-effective way for low-cost devices with limited battery/power capabilities to form an ad-hoc network. Given the lack of Power resources, the throughput is low and does not conform to many QoS requirements. The LR-WPAN details are published by the IEEE 802.15.4 standard and the strength of the protocol lies in its simplicity and flexibility (https://standards.ieee.org/standard/802_15_4-2011.html).

The protocol only defines the Physical and the MAC layer details and allows other protocols to operate on top of the MAC layer. The devices in an LR-WPAN network can operate either as a Full-Functional Device (FFD) or a Reduced-Function Device (RFD), Figure 4. An FFD can act as a PAN coordinator which roughly can be translated to as the Master device in a Master-Slave framework. An RFD, on the other hand, is a pure slave node. Such devices are intended for very simple tasks (e.g., turning on a light switch). An FFD device can handle a higher level of traffics and hence are characterized by stronger resources compared to an RFD.

The task of the Physical layer (PHY) is similar to the one in OSI model i.e. physically manage the physical layer to send the frames. In this case, the PHY consists of managing the radio signals. The MAC layer is responsible for channeling the packets from higher layers to the PHY layer.

The devices can operate in one of two network topologies namely 1) Star topology and 2) Mesh topology. In a star network topology, devices choose a PAN coordinator whose job is to disseminate information to other devices. The star topology depicted in Figure below follows a typical Master-Slave model.

The task of the Physical layer (PHY) is similar to the one in OSI model i.e. physically manage the physical layer to send the frames. In this case, the PHY consists of managing the radio signals. The MAC layer is responsible for channeling the packets from higher layers to the PHY layer.

The devices can operate in one of two network topologies namely 1) Star topology and 2) Mesh topology. In a star network topology, devices choose a PAN coordinator whose job is to disseminate information to other devices. The star topology depicted in Figure below follows a typical Master-Slave model.

A mesh topology, on the other hand, follows a peer-to-peer model. While the presence of PAN coordinator is still mandated, the nodes can contact each other directly as shown in Figure 5. Note that unlike other mesh topologies, the PAN coordinator is not responsible for routing the messages to other devices but rather the devices can connect to each other directly.

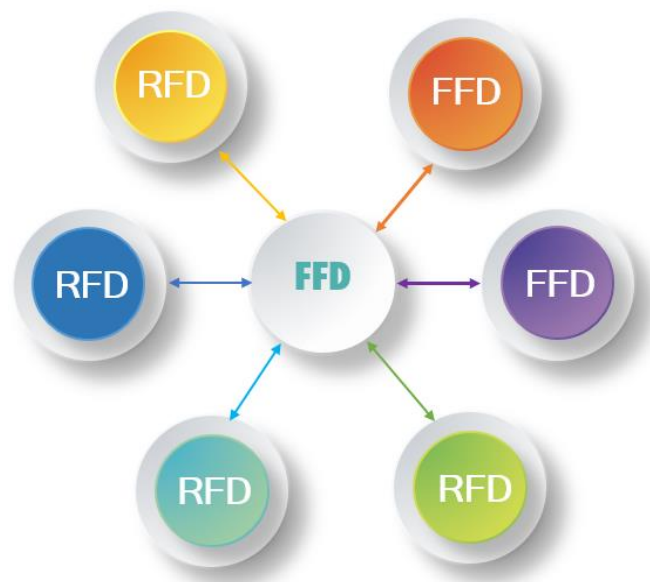

**Figure 3** Star Topology

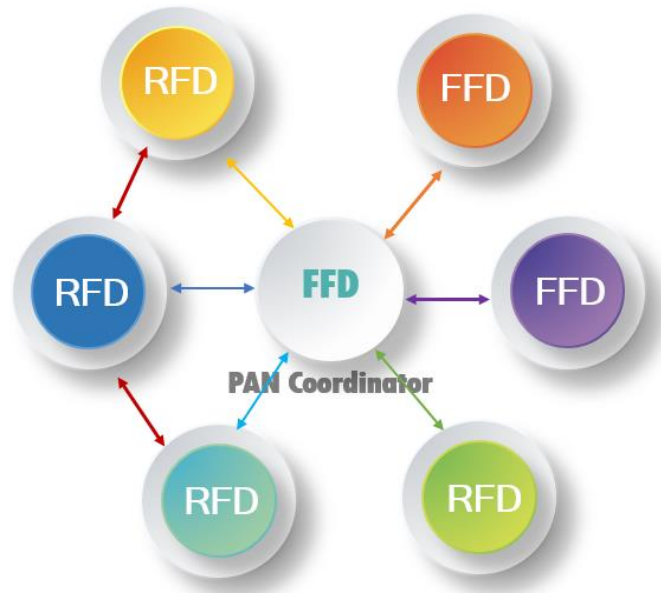

**Figure 4** Mesh Topology

The PAN coordinator is responsible for assigning unique short-term addresses to the devices operating within its own network. The devices can be part of more than one PAN network as each network operates on its own unique frequency. A Star network is suitable for a home or industrial control environment. A mesh or peer-to-peer network provides the flexibility to span wider areas and applications.

As opposed to the Bluetooth, the 802.15.4 does not provide the complete OSI model support. Rather it is the platform upon which other protocols such as 6LoWPAN, Zigbee, WirelessHART and ISA100.11a are built. The discussion of these protocols is beyond the scope of this chapter.

### 2.2.3 High Rate WPAN

A high Rate WPAN (HR-WPAN) – defined by the IEE standard 802.15.3 – is an attempt to provide a similar capability to the LR-WPAN but increase the data transfer rate. The standard utilizes the millimeter-wave-based alternate physical layer. This allows the throughput to increase anywhere from 11 Mbps to 55 Mbps (https://standards.ieee.org/standard/802_15_3-2016.html). Such technology is imperative in providing streaming abilities such as live video feed from the source to destination spanning huge distances.

While the idea is similar to 802.15.4, the HR-WPAN builds upon the concept of Piconets. The protocol defines the Protocol Adaptation Layer (PAL) in addition to the Physical and MAC layers as shown in Figure 6. The PAL Layer allows the higher level protocols to communicate with the MAC and in turn the physical layer of the HR-WPAN.

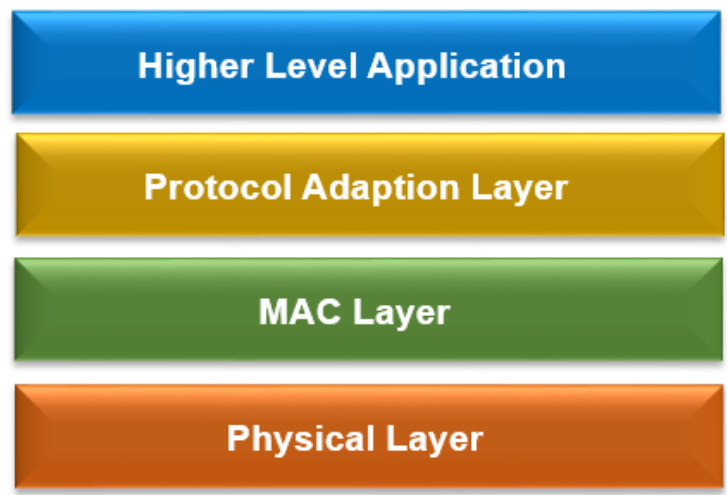

**Figure 5** 802.15.4 Protocol Stack

Once the protocol layer is established, the protocol requires a Piconet Coordinator (PNC) similar to the PAN coordinator in LR-WPAN. The PNC coordinator finds a channel through which it can communicate to the devices that are in its span. In addition to the Piconet capability discussed in the Bluetooth section, the standard allows for child Piconets and neighbor Piconets in case a particular Piconet cannot find a frequency channel to communicate. Both the child and neighbor Piconets can help the extend the range of a given Piconet.

The child Piconets use the same channel to communicate with devices in its own network or the parent network. Given that the frequency is the same, the networks utilize something called Channel Time Allocation (CTA) with the help of a scheduler.

A neighbor Piconet on the other hand requests from a PNC to share the frequency to communicate with its members using the same frequency utilizing the CTA. As opposed to the child Piconet, a neighbor Piconet cannot communicate directly to the devices under the control of the PNC of the neighbor PNC.

If a PNC wishes to exit the Piconet, it will try to find a suitable device which can take over the PNC responsibilities. In case no such device exists, the PNC will send a message to all the devices informing them about the termination of Piconet. In case a child or neighbor Piconet exist, they will continue to operate without hindrance as the parent/neighbor channel frequency will be available to them.

### 2.2.4 Body Area Networks

The IEEE standard 802.15.6 defines the Body Area Networks (BANs). Given that the standard still falls under the 802.15 (Personal Area Networks), it is safe to conclude that the concepts used in the other standards are in play. However, there is one more element that the standard introduces – Human Body Communications (HBC) that occurs at the physical layer. The communication happens using the Electric Field Communications (EFC) centered at 21MHZ. This layer allows the human body to send signals that are then converted to the MAC layer.

Once the above happens, the next question would be regarding the topology the network uses. As opposed to the LR-WPAN and HR-WPAN, the only network topology BANs use is that of Star topology. Thus there needs to be a device that acts as the coordinator – called the Hub. The remaining devices connect to the Hub (The Hub-Spoke model is used synonymously with Master-Slave model).

The communication between the hub and the devices happens in terms of exchanging frames. Given the sensitivity of the information, the standard defines three security levels as follows:

Level 0 – Unsecured communication

Level 1 – Authentication

Level 2 – Authentication and Encryption

If the hub intends to communicate with the nodes in level 1 or level 2 mode, the nodes can be in any one of the following states.

1. Orphan – The device is not connected to the hub
2. Connected – The device and hub are allowed to exchange secure frames with each other but not unsecured frames
3. Associated – The device can transfer only information regarding connection and being in a secure state
4. Secured – The device and hub can exchange information securely with each other.

On the other hand, insecure communication allows the nodes to be in one of the two following states.

1. Orphan – The device is not connected to the hub
2. Connected – The device and hub are allowed to exchange secure frames with each other but not unsecured frames

# 3 Prevalent IoMT Applications

## 3.1 IoMT services and applications

Two aspects of the IoMT can be found in the literature namely the services such as wearable devices and the applications such as ECG or blood pressure monitoring (Gatouillat et al., 2018; Magsi et al., 2018; UST, 2017; Islam et al., 2015). Figure 7 illustrates an example of the main services and application in an IoMT. By examination of this figure, it is clear that services are used to develop IoMT application while applications are directly used by patients. Aside from their utility in supervising and managing daily health and normal well-being, IoMT devices have additionally been utilized for chronic disease management and prevention, remote assistant living and intervention, improved drug management, and wellness and preventive care in any remote location (Shelke, 2018; Joyia, 2017).

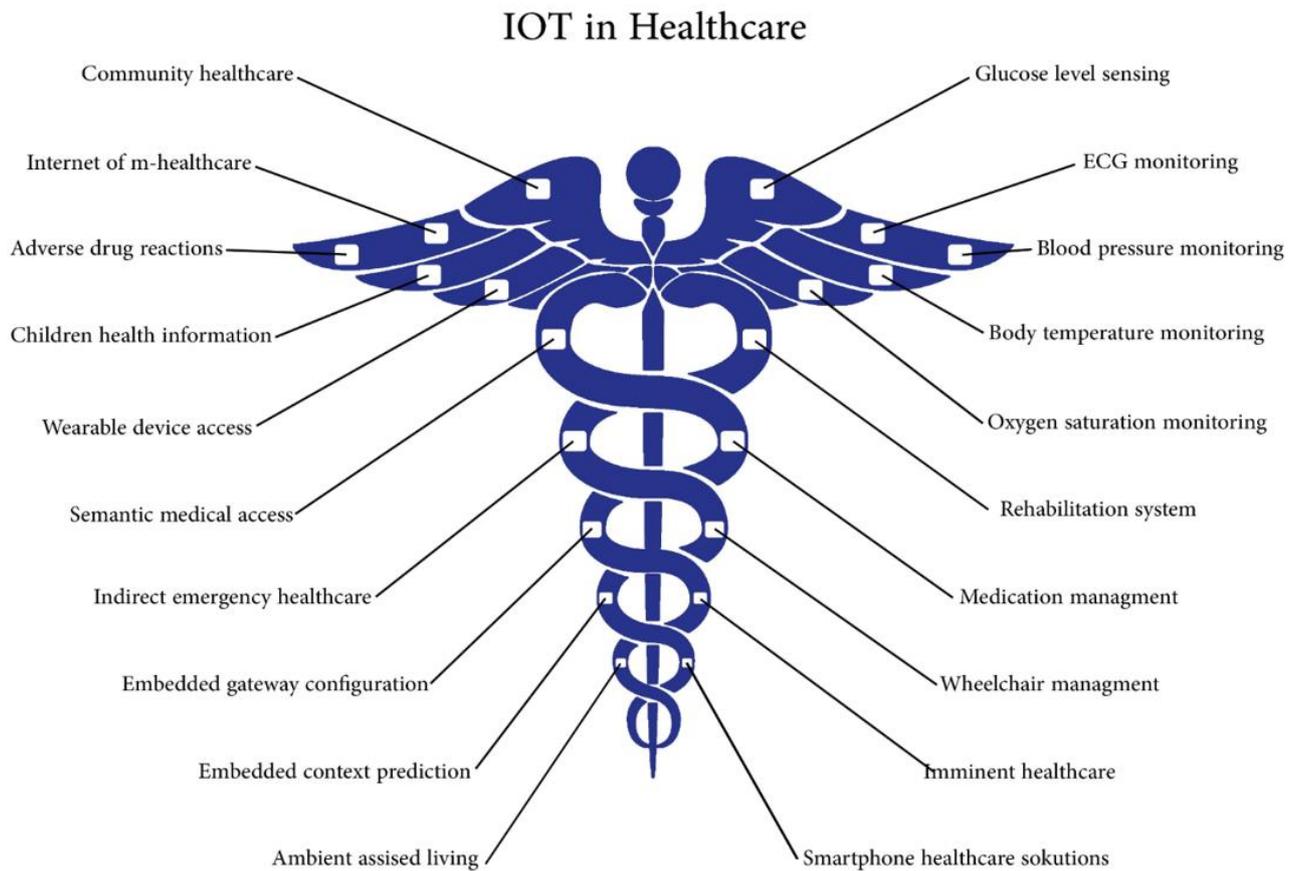

**Figure 6: Main services and application in an IoMT**

Different wearable devices exist in the market such activity monitors, automated external defibrillator, blood pressure monitor, blood glucose meter, fall detector, fitness and heart rate monitor, multi-parameter monitor, programmable syringe pump, pulse oximeter, smart pill dispenser, smart watches, and wearable injector, to cite only few (Krohn et al., 2017; Cruz et al., 2018, Bayo-Monton et al., 2018,

Metcalf et al., 2016a, Metcalf et al., 2016b). The application domain of IoMT may include Chronic Diseases (CD), Health Fitness Management (HFM), Home-based Medical Health (HMH), Hospital Monitoring (HM), Human Activity Recognition (HAR), Medical Nursing (MN), Patient Physiological Conditions (PPC), Patients' Wearable IoMT Devices, Pediatric and Elderly Care (PEC), Remote Patient Monitoring (RPM), Simultaneous Reporting and Monitoring (SRM), and Tracking Patient Medication (TPM).

Besides these application domains, Telemedicine Monitoring Remote Consultation (TMRC) represents a new approach to medicine which can be defined as the utilization of medical history and information shared from one party to another via electronic communications to enhance, assist or maintain patients' health status (Field et al., 2002; Xiao, 2008; Giorgio, 2011, Shams, 2014). For a border definition, telemedicine is associated with the term 'Telehealth" which defines the remote healthcare (Lazarev, 2016; Higgs, 2014). There are many useful applications used to ease the medical process under this approach. According to a recent study and analysis by Deloitte Centre for Health Solutions (Deliotte, 2018), the market for connected medical devices is predicted to grow from $14.9 billion in 2017 to $52.2 billion in 2022. Today, there are more than 500,000 different types of medical devices including wearable external medical devices, implanted medical devices, and stationary medical device as reported by Deliotte (2018). Figure 8 shows the trend of the global wearable computing devices from 2017 to 2019 based on the data published by ABI Research (ABI research, 2018; Dias et al, 2018). According to this data, healthcare wearable devices show an increasing growth and even exceed wearable devices used in sport activities. In another study, the "Wearable Medical Devices - Global Market Outlook (2017-2026)" (Global Market Outlook, 2018) reported that in 2017 global wearable medical devices accounted for $6.05 billion, this trends will continue and it is expected to reach $29.53 billion in 2026.

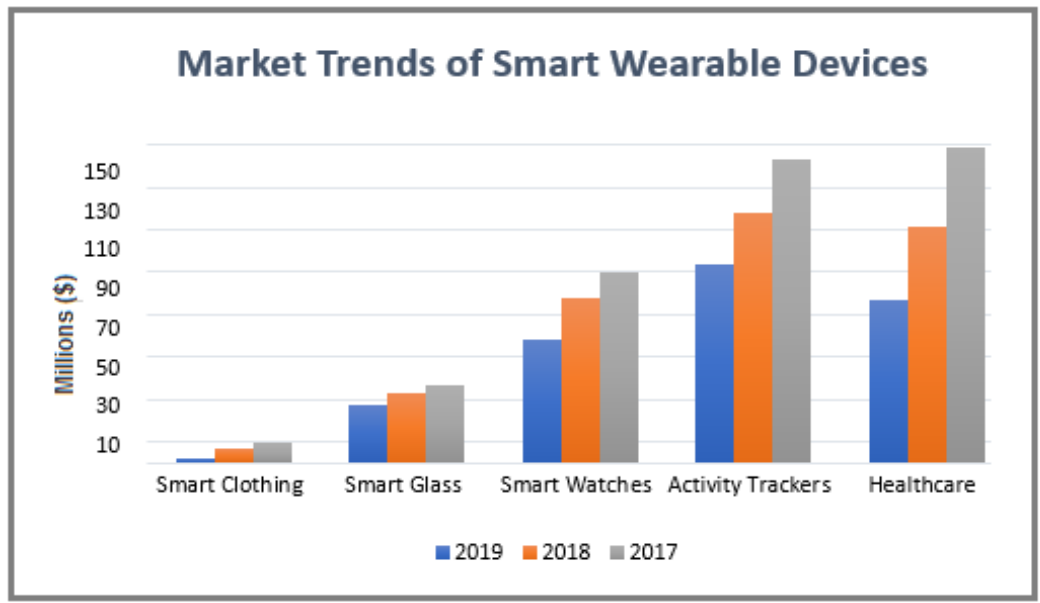

**Figure 7  Trends of the global market value of wearable devices, 2017-1019. Adapted from ABI Research (2014)**

Wearable IoMT devices are smart devices that can deliver efficient personal experience and services, and they are the heart of smart IoT healthcare solutions. These wearable devices, integrated with telemedicine, are used in continuous patient monitoring. They integrate into applications including measuring vital signs and exchange reliable and secure information through the IoT. The field of wearable health monitoring systems is advancing in minimizing the size of wearable devices and medical sensors (Haghi et al., 2017). Another application related to healthcare is the Human Activity Recognition (HAR) where IoT is used to remotely monitor vital human signs (Rodriguez et al., 2017). Human activity detection and analyzing is a challenging task because it requires considerable time-consuming and high-cost hardware. Accurate recognition of human activities could help in healthcare services for improved patient recovery training guidance, or an early indicator of emergency medical conditions that elder patients may encounter, such as falls and heart failures (Liu et al., 2016).

In the case of tele patients with chronic diseases, a home monitoring system is used to monitor multiple features such as weight scales, pulse oximeters, glucometers, and blood pressure cuffs. Readings are recorded in personal health records, and warnings are sent wirelessly to health-care providers when readings fall beyond the normal range. Diabetes is one of the common chronic diseases in which there are high blood glucose levels over a lengthy period. Managing this requires non-invasive glucose sensors to monitor the level patterns. An IoMT method to manage this is to have sensors from patients connected through IPv6 connectivity to healthcare providers. This device consists of a blood glucose collector, a mobile phone, and a background processor (Islam et al., 2015). Another crucial case is patients with heart disease, monitoring the electrical activity of the heart is an essential procedure for these patients. The early diagnosis of the arrhythmias can prevent a major risk. To prevent that, associating IoT in ECG monitoring give the potential to offer accurate features and information about the patient's signs remotely (Islam et al., 2015). For monitoring body temperature, the IoMT promises to develop practical solutions to the health care services by allowing monitoring temperature as it is a vital body parameter. The involvement of IoMT has integrated temperature sensor interface with wireless media. The sensor in smart temperature patch can detect the data and send it over Bluetooth or Wi-Fi connectivity to a specific cloud application where the concerned doctor can perform data-analysis for the patient.

For elderly care, it is important to note that supporting the independent life of elderly people in their living place safely can make them more confident by ensuring better autonomy and giving them real assistance. This is the main objective of the Ambient Assisted Living (AAL), an IoT platform powered by artificial intelligence that uses the information and communication technologies to serve this objective. Activity recognition and behavior understanding are the desired results of using a variety of sensors in AAL (García, et al., 2017; Bevilacqua et al., 2014; Monekosso et al., 2015). This can be direct via wearable sensors or indirectly through environmental sensors and stream analysis. On tracking patient medication, it is becoming possible for patients to track their medication ingestion. Figure 9 illustrates a typical Wireless Body Sensor Network (WBSN) where the information is received from multiple body sensors and transmitted remotely via a gateway to the central server for storage or decision. It is important; however, to danger with the noncompliance with medication schedule which can lead to serious complications for the patient's health (Kuzela, 2015). The revolutionary of intelligent smart pill technology provided a comprehensive solution to the patient's noncompliance in medications. The pill bottle, FDA approved (FDA, 2017), is embedded with an Ingestible Event Marker (IEM) sensor that tracks the medication intake and reminds patients to take their medication at appropriate times. Once the patient swallows it, the IEM can detect the ingestion and communicates this data to a wearable sensor or mobile device through cloud infrastructure (Demeo et al., 2014, Botta et al., 2014, Botta et al., 2016).

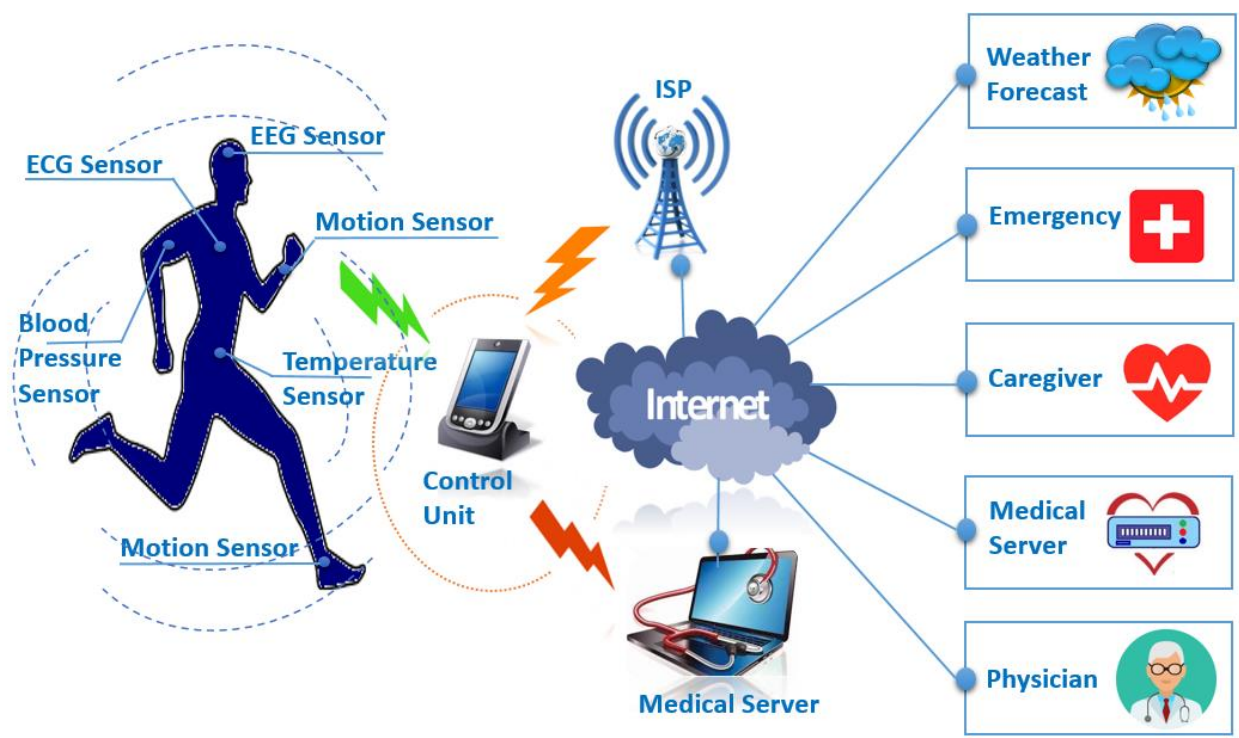

**Figure 8** A typical Wireless Body Sensor Network (WBSN)

**3.2  IoMT Companies Leading the Way**

With the overwhelming growth of healthcare electronic devices connected to each other to send and receive safely and effectively sensitive data and information, the Food and Drug Administration (FDA, 2017) has examined cyber-security risks and released on Sept 6, 2017, its final guidance on "Design Considerations and Premarket Submission Recommendations for Interoperable Medical Device". FDA identified six issues that medical device manufacturers should take into account: (1) identify the purpose of the electronic interface including the type and data exchanged, (2) determine the anticipated user, (3) consider risk management including risk that may arise from other users connecting through the interface, (4) verification and validation by maintaining and implementing appropriate verification and validation of the device functionality , (5) labelling considerations, and (6) use of consensus standards related to the interoperability of the medical device.

The key players in the IoMT market include Adheretech (https://www.adheretech.com/) patient support programs with its smart pill bottle and underlying software to detect compliance with medications and alerts patients in case they forget or miss a dose, or detect a serious problem in which case the patients' pharmacy is notified. AliveCor (https://www.alivecor.com) records patient EKG using KardiaMobile Smartphone. Bosch Healthcare (https://www.bosch-healthcare.com) developed a breath analysis device for asthma patients. Capsule Technologies (https://www.capsuletech.com) takes control of patient medical device data and provide powerful data configuration options. Cerner Corporation (https://cerner.com/), a supplier of health information technology (HIT) ranging from medical devices to electronic health records (EHR) to hardware. Cisco Inc. (https://www.cisco.com) support critical applications such as imaging and electronic medical records. DeepMind Health (https://deepmind.com), helps clinicians get patients from test to treatment. Diabetizer Ltd. (https://diabetizer.com), as indicated by its name, analyzes diabetes and generates an assessment of patient current health status. Ericsson, with the possibility of study people's behaviors and values, GE Healthcare (https://www.gehealthcare.com), a leading provider of medical imaging, monitoring, biomanufacturing, and cell and gene therapy technologies. Honeywell Life Care Solutions provides a remote patient monitoring solution. IBM Watson for drug discovery, Medtronic Inc. (www.medtronic.com), a global leader in medical technology, services, and solutions. Microsoft (https://www.microsoft.com), allow patients to receive care at home, monitor medical assets, and track equipment usage. Proteus Digital Health (https://www.proteus.com/) measures the effectiveness of the medication treatment as well as helping physicians to improve clinical outcomes. Qualcomm Life Inc. (https://www.qualcomm.com) Integrates health care data for access anywhere, anytime. Stanley Healthcare (https://www.stanleyhealthcare.com/) provide analytics solutions to ensure safety and security for senior living organizations, hospitals, and health systems. Vodafone integrates patient healthcare devices with their hardware, software and manages their connectivity. Zebra Technologies (https://www.zebra.com) connects medical providers with patient records for better care and better outcomes.

# 4 Conclusions and Future Directions

IoMT offers huge potential and a great set of challenges to researchers. PANs form the underlying basis of IOT and IOMT. In this chapter, we have focused on four standards namely Bluetooth, IEEE 802.15.3 and 802.15.4. Furthermore, we have also briefly discussed Body Area Networks (BANs) along with some of the prevalent IoMT applications and the vendors that provide such services. Any disruption in PANs would render the entire service unusable and hence deserves a lot of attention from researchers. Our work aims at introducing the readers to the intricacies of PANs and their importance to Health Informatics. PANs are characterized by very limited storage and hence rely heavily on transferring data to stable persistent storage via the underlying networking technologies. The chapter also introduces the reader to various types of networks that are in use and how the choice affects the type of applications built upon such networks.

## 4.1 Future Research Directions

PANs have limitations both in terms of functionality and the security aspect. The protocols used are relatively new and hence future work should look into the strengths of weaknesses of such protocols. In addition, the coexistence of such protocols presents an interoperability challenge that also needs to be explored.

Furthermore, researchers will also need to classify various applications in terms of the nature of the temporal aspects of the application being studied. Real-time applications have high throughput requirements and certain underlying networks will be unsuited for certain applications.

Lastly, the researchers need to look into the security and privacy aspect of such networks. Given the federal laws governing the privacy of patients' data, it remains to be seen the implications of widespread adoption of such technologies.

## 4.2 Recommended Assignments

- What kind of security threats exist in terms of PANs?
- How the performance of BANs is affected by the security and encryption of packets?
- Can Bluetooth standard fulfill the security and performance requirements eliminating the need of Low Rate and High Rate WPANs?